\pdfoutput=1
\documentclass[submission,copyright,creativecommons]{eptcs}
\providecommand{\event}{NCMA 2026} 

\usepackage{iftex}

\ifpdf
  \usepackage{underscore}         
  \usepackage[T1]{fontenc}        
\else
  \usepackage{breakurl}           
\fi

\usepackage{amsmath,amsgen,latexsym}
\usepackage{amstext,amssymb,amsfonts,latexsym}
\usepackage{theorem}
\usepackage{pifont}
\usepackage{graphics}

 \newcommand{\n}{\noindent}
 \newcommand{\s}{\smallskip}
 \newcommand{\hs}[1]{\hspace*{ #1 mm}}
 \newcommand{\vs}[1]{\vspace*{ #1 mm}}

 \newcommand{\nat}{\mathbb{N}}
 
 \newcommand{\integer}{\mathbb{Z}}

 \newcommand{\co}{\mathrm{co}\mbox{-}}

 \newcommand{\CC}{{\cal C}}

 \newcommand{\KK}{{\cal K}}
 \newcommand{\LL}{{\cal L}}

 \newcommand{\MM}{{\cal M}}

 \newcommand{\PP}{{\cal P}}

 \newcommand{\dl}{\mathrm{L}}
 \newcommand{\nl}{\mathrm{NL}}
 \newcommand{\p}{\mathrm{P}}

 \newcommand{\poly}{\mathrm{poly}}

 \newcommand{\cfl}{\mathrm{CFL}}

\theoremstyle{plain}
\theoremheaderfont{\bfseries}
 \newtheorem{theorem}{Theorem}[section]
 \newtheorem{lemma}[theorem]{Lemma}
 \newtheorem{proposition}[theorem]{{\bf Proposition}}
 \newtheorem{corollary}[theorem]{Corollary}

 {\theorembodyfont{\rmfamily}  }
 {\theorembodyfont{\rmfamily} }
 {\theorembodyfont{\rmfamily} }

 \newenvironment{proofsketch}{\par \noindent
            {\bf Proof Sketch. \hs{2}}}{\hfill$\Box$ \vspace*{3mm}}

 \newenvironment{proofsketchof}[1]{\vspace*{5mm} \par \noindent
         {\bf Proof Sketch of #1.\hs{2}}}{\hfill$\Box$ \vspace*{3mm}}

 \newenvironment{yproof}{\par \noindent
            {\bf Proof. \hs{2}}}{\hfill$\Box$ \vspace*{3mm}}

 \newcommand{\floors}[1]{\lfloor #1 \rfloor}
 \newcommand{\pair}[1]{\langle #1 \rangle}

\newcommand{\ignore}[1]{}

 \newcommand{\oned}{1\mathrm{D}}

 \newcommand{\onep}{1\mathrm{P}}
 \newcommand{\onebp}{1\mathrm{BP}}
 \newcommand{\onen}{1\mathrm{N}}
  \newcommand{\twou}{2\mathrm{U}}
 \newcommand{\twod}{2\mathrm{D}}
 
 \newcommand{\twon}{2\mathrm{N}}

 \newcommand{\twobp}{2\mathrm{BP}}

 \newcommand{\oneu}{1\mathrm{U}}

 \newcommand{\twoupd}{2\mathrm{UPD}}
 \newcommand{\onedpd}{\mathrm{1DPD}}
 \newcommand{\onenpd}{\mathrm{1NPD}}
 \newcommand{\twodpd}{\mathrm{2DPD}}
 \newcommand{\twonpd}{\mathrm{2NPD}}
 \newcommand{\logdcfl}{\mathrm{LOGDCFL}}
 \newcommand{\logcfl}{\mathrm{LOGCFL}}

 \newcommand{\twof}{2\mathrm{F}}
 \newcommand{\twofct}{2\mathrm{FCT}}
 \newcommand{\lintwod}{\mathrm{lin2D}}

 \newcommand{\linonep}{\mathrm{lin1P}}
 \newcommand{\linoner}{\mathrm{lin1R}}

 \newcommand{\onedct}{\mathrm{1DCT}}
 \newcommand{\onenct}{\mathrm{1NCT}}
 \newcommand{\twodct}{\mathrm{2DCT}}
 \newcommand{\twonct}{\mathrm{2NCT}}
 \newcommand{\twouct}{\mathrm{2UCT}}

 \newcommand{\onedpdct}{\mathrm{1DPDCT}}
 \newcommand{\onenpdct}{\mathrm{1NPDCT}}
 \newcommand{\twodpdct}{\mathrm{2DPDCT}}
 \newcommand{\twonpdct}{\mathrm{2NPDCT}}

 \newcommand{\ptime}{\mathrm{ptime}\mbox{-}}
 \newcommand{\etime}{\mathrm{etime}\mbox{-}}
 \newcommand{\polylog}{\mathrm{polylog}}

 \newcommand{\oner}{1\mathrm{R}}

\newcommand{\nauxpdaspti}[2]{\mathrm{NAuxPDA},\!\mathrm{SPTI}( #1, #2 )}
\newcommand{\dauxpdaspti}[2]{\mathrm{DAuxPDA},\!\mathrm{SPTI}( #1, #2 )}

\begin{document}


\title{How Can Size and Ceiling Bounds Affect the Complexity of Nonuniform Automata Families?}
\author{Tomoyuki Yamakami
\institute{Faculty of Engineering, University of Fukui, 3-9-1 Bunkyo, Fukui 910-8507,  Japan}
\email{TomoyukiYamakami@gmail.com}
}
\def\titlerunning{Complexity of Nonuniform Automata Families}
\def\authorrunning{T. Yamakami}

\maketitle

\begin{abstract}
In the past literature, families of two-way finite automata and pushdown automata having limited state complexity (i.e., the total number of inner states) and stack-state complexity (i.e., the total number of inner states multiplied by the total number of strings ``pushable'' to a stack), have been studied in direct connection to (mainstream)  space-bounded complexity classes equipped with Karp-Lipton style advice of limited size when all inputs given to the automata have bounded length. Here, we acknowledge two major factors---size and ceiling---of such families, which have a significant impact on the complexity of finite and pushdown automata families, where the ``size'' refers to (stack-)state complexity and the ``ceiling'' refers to an input's length bound. In this line of study, we further explore those effects caused by different sizes and ceilings.

\s

\n{\bf Keywords.} nonuniform state complexity, promise problem,  stack-state complexity, size, ceiling, advice
\end{abstract}


\sloppy
\section{Background and Challenges of This Work}\label{sec:introduction}


In 1959, Rabin and Scott \cite{RS59} published a paper claiming that, for each fixed positive integer $n$,  each one-way nondeterministic finite automaton (or 1nfa, for short) $N_n$ of $n$ inner states can be converted into another computationally equivalent\footnote{Two machines are \emph{computationally equivalent} if their outcomes coincide on every input.}
one-way deterministic finite automaton (or 1dfa) $M_n$ of $2^{n}$ inner states.
The total number of inner states used to describe each finite automaton, known as the \emph{state complexity}, has served as a useful complexity measure indicating the ``size'' of the finite automaton. In other words, the result of Rabin and Scott asserts that every 1nfa of size $n$ can be simulated on an appropriate 1dfa of size at most $2^n$.  After the publication of their paper, researchers have been wondering if the value $2^n$ can be significantly reduced.

As for the simulation of an $n$-size 2nfa $N_n$ on an $n^{O(1)}$-size 2dfa $M_n$,  where 2nfa and 2dfa are respectively the two-way versions of 1nfa and 1dfa, Berman and Lingus \cite{BL77} reported an intimate connection to the $\dl=?\nl$ question, where  $\dl$ is the deterministic log-space complexity class and $\nl$ is its nondeterministic variant.

In 1978, Sakoda and Sipser \cite{SS78} studied collectively (nonuniform) families of finite automata $M_n$ of  $n^{O(1)}$  size (i.e., polynomial size) in order to solve given families of promise (decision)  problems\footnote{A \emph{promise (decision) problem} is a pair of disjoint
sets over the same alphabet.} $(L_n^{(+)},L_n^{(-)})$ indexed by natural numbers $n$.
As a concrete example, assuming a suitable binary encoding $\pair{G}$ of a graph $G$, let us consider the family $\LL_{bipartite} = \{(L_n^{(+)},L_n^{(-)})\}_{n\in\nat}$ of promise problems defined by $L_n^{(+)}=\{x\in\{0,1\}^*\mid |x|\leq 2^n, x=\pair{G}, \text{ $G$ is a bipartite graph}\}$ and $L_n^{(-)}=\{0,1\}^* - L_n^{(+)}$ for each index $n$.   This family $\LL_{bipartite}$ can be solved by a suitable family $\{N_n\}_{n\in\nat}$ of polynomial-size 2nfa's.
A series of followup studies has been conducted intensively in the past literature
\cite{Gef12,Kap09,Kap12,Kap14,KP15,Yam18,Yam19a,Yam19b,Yam21,Yam23}.

The focal object of Sakoda and Sipser is actually the nonuniform families of ``polynomial-size'' finite automata. Here, the ``size'' describing a finite automaton signifies the computational complexity  of this automaton, more akin to the (work) space complexity measure of a Turing machine equipped with an external information source known as ``advice''.
This association contributes, as a hidden gem, to the study of automata theory by way of (mainstream) computational complexity theory.
As for pushdown automata families, in contrast, the ``size'' refers to its \emph{stack-state complexity}, which indicates the total number of inner states multiplied by the total number of ``pushable'' strings \cite{Yam21,Yam23} because stack symbols are translatable in some sense to inner states.

The study of those ``families'' of finite and pushdown automata takes a similar advantage to that of (nonuniform)  Boolean circuit families; however, a major difference also lies in the point that, whereas  each circuit appearing in a Boolean circuit family is limited to particular input size, each finite automaton in an automata family is not  in general, as the previous example $\LL_{bipartite}$ shows.
Therefore, it is sometimes beneficial to cap an appropriate upper bound on the length of input strings fed into underlying finite and pushdown automata. Such a cap is known as a ``ceiling'' and has played a key role
in \cite{BL77,Kap14,SS78}, in which, for instance, the standard log-space complexity classes, $\dl$  and $\nl$, supplemented by polynomial-size advice can be expressed  in terms of collections of finite automata families
when input strings are all limited to having \emph{polynomial ceiling} (i.e., polynomially bounded length).

Through those studies, there are, in essence, two major
factors---\emph{size} and \emph{ceiling}---which significantly affect the overall performance of each automata family. Those factors may directly relate to the complexity measures of work space and advice size used in standard complexity theory.
The purpose of this work is therefore to further explore how such factors contribute to the shift of the overall complexity of nonuniform automata families.

Associated with those factors, the aforementioned initial studies of automata families have been expanded into further directions.
As natural extensions of polynomial-size automata families, for instance,
Kapoutsis \cite{Kap09,Kap12} studied the behaviors of  finite automata families of \emph{exponential (i.e.,  $2^{n^{O(1)}}$) size}.
As for ceiling bounds, Kapoutsis also considered promise problem families of \emph{superpolynomial ceilings}.   When ceiling bounds of polynomial-size automata families are expanded from polynomials to, e.g.,  exponentials, they in turn relate to loglog-space complexity classes supported by  polylog-size advice \cite{Kap14}.


Unfortunately, we still lack a whole picture depicting the effects of the combinations of sizes and ceilings for nonuniform families of finite and pushdown automata.
The rest of this work is organized as follows. The basic notions and various notations will be explained in Section \ref{sec:preparation}.
The effects of polynomial ceilings will be discussed in Section \ref{sec:poly-ceilings} and those of exponential ceilings will be studied in Section \ref{sec:exp-ceilings}. From additional viewpoints, polynomial-size and superpolynomial-size automata will be examined separately in Sections \ref{sec:poly-size-automata} and \ref{sec:exp-size-automata} as well as Section \ref{sec:exp-ceilings}.

All omitted proofs will be included in a forthcoming complete version of this paper.

\section{Foundations of This Work}\label{sec:preparation}

We will explain the basic notions and notation that the reader needs to read through the rest of this work.

\subsection{Sets, Numbers, and Alphabets}

We use the standard notations, including $\nat$ (natural number class with $0$). For positive integers, we write $\nat^{+}$ for $\nat-\{0\}$. For two integers $m,n$ with $m\leq n$, $[m,n]_{\integer}$ denotes the \emph{integer interval} composed of all integers between $m$ and $n$ in comparison with the \emph{real interval}  $[r_1,r_2]$. We often abbreviate as $[n]$ the integer interval $[1,n]_{\integer}$ for any number $n\in\nat^{+}$.
In this work, \emph{polynomials} take nonnegative integer coefficients and \emph{logarithms} are all  taken to the base $2$. For convenience, we further set $\log{0}$ to be $0$.
Given a set $Q$, $\PP(Q)$ denotes the \emph{power set} of $Q$.

The notation $\lambda$ denotes the \emph{empty string} of length $0$.
In general, a \emph{promise (decision) problem over an alphabet $\Sigma$} is a pair $(L^{(+)},L^{(-)})$ of disjoint subsets of $\Sigma^*$. In particular, whenever $L^{(+)}\cup L^{(-)}=\Sigma^*$, $L^{(+)}$ is simply called a \emph{language} and $L^{(-)}$ is said to be the \emph{complement} of $L^{(+)}$. Similarly, $(L_n^{(-)},L_n^{(+)})$ is the \emph{complement} of $(L_n^{(+)},L_n^{(-)})$ and denoted $\co(L_n^{(+)},L_n^{(-)})$.

A function $f$ from $\nat$ to $\nat$ is called \emph{polynomially bounded} (or p-bounded, for short) if there exists a polynomial $p$ such that $f(n)\leq p(n)$ holds for all $n\in\nat$. In contrast, $f$ is said to be \emph{polynomially honest} (or p-honest) if there is a polynomial $p$ satisfying  $n\leq p(f(n))$  for all $n\in\nat$. The composition $f\circ g$ of two functions $f$ and $g$ is defined
by $f\circ g(x) = f(g(x))$ for all inputs $x$.

A \emph{partial function} from $\Sigma^*$ to $\Gamma^*$ for two alphabets $\Sigma$ and $\Gamma$ is expressed in this work as a pair $(f,D)$ with  a set $D$ of valid (or defined) inputs $x$, which means that the outcome of $f$ on $x$, denoted $f(x)$, is defined.

\subsection{Finite Automata, Pushdown Automata, and Finite Transducers}\label{sec:pushdown-automata}

As a foundation, this work uses the following machine models: \emph{one-way deterministic finite automata} (or 1dfa's, for short), \emph{one-way nondeterministic finite automata} (or 1nfa's),  \emph{one-way deterministic pushdown automata} (or 1dpda's), \emph{one-way nondeterministic pushdown automata} (or 1npda's), and their two-way variants
(or 2dfa's, 2nfa's, 2dpda's, and 2npda's). The acceptance/rejection of computation paths of a machine is determined by simply entering accepting/rejecting inner states of the machine.

We remark that every pushdown automaton is allowed to make its input-tape head stay still (known as a \emph{$\lambda$-move}).
Moreover, in general, the two-way machines
allow their input-tape heads to move in all directions including $\lambda$-moves. By contrast, the one-way finite machines must move their tape heads to the right \emph{at every step}.
Notice that each input is always marked by two endmarkers $\rhd$ and $\lhd$. 
Given a machine, say, $M$, we say that $M$ \emph{solves} a promise problem  $(L^{(+)},L^{(-)})$ if   $M$ accepts all inputs $x$ in $L^{(+)}$ and  $M$ rejects all inputs $x$ in $L^{(-)}$.

For simplicity, we also view 1nfa's as \emph{one-way probabilistic finite automata} (or 1pfa's) by assigning equal ``probability'' to their nondeterministic choices made at every step. Here, we remark that 1pfa's were generally defined by stochastic matrices. See, e.g., \cite{Yam23} for those matrix-based definitions of 1pfa's.  We say that a 1pfa $M$ recognizes a language $L$ with \emph{unbounded-error probability} if, for all strings $x\in L$, $M$ accepts $x$ with probability more than $1/2$ and, for all $x\notin L$, $M$ rejects $x$ with probability at least $1/2$.
Moreover, $M$ recognizes $L$ with \emph{one-sided-error probability} if, for all $x\in L$, $M$ accepts $x$ with probability more than $1/2$ and, for all $x\notin L$, $M$ rejects $x$ with probability $1$.

Given two alphabets $\Sigma$ and $\Gamma$, a \emph{two-way deterministic finite transducer} (or a 2dft) is a 2dfa equipped with a write-once\footnote{A tape is said to be \emph{write once} if its tape head never moves to the left and, whenever the tape head writes non-blank symbol, it should move to the next blank cell.} output tape, where it writes at most one symbol on the output tape at every step. This output restriction is slightly different from that of its one-way version (or 1dft) given in \cite{Yam23}.


It is possible to significantly reduce the number of nondeterministic choices made at every step at the cost of increasing the state complexity. Given an arbitrary integer $k\geq2$, we say that a 2nfa $N$ is of \emph{$k$-choice} if $N$ makes at most $k$ nondeterministic choices at every step.

\begin{lemma}\label{two-choice-change}
Given an $n$-size 2nfa $N$, there exists a computationally equivalent 2-choice 2nfa $N'$ of size at most $5n^2$.
\end{lemma}


The \emph{push size} $e$ of  a pushdown automaton $M$ is the maximum value $|w|$ in the set $\{|w|: (p,w)\in\delta(q,\sigma,a)\}$, where $\delta$ is a transition function of $M$. The \emph{stack-state complexity} of $M$ is $|Q||\Gamma^{\leq e}|$ \cite{Yam20,Yam21,Yam24,Yam24b} in comparison with the \emph{state complexity} (i.e., $|Q|$) of a finite automaton, where $Q$ is a set of inner states and $\Gamma$ is a stack alphabet of $M$.

Two machines $M$ and $N$ over the same input alphabet are \emph{computationally equivalent} if the outcomes (i.e., acceptance or rejection) of $M$ and $N$ agree on all inputs.


A \emph{(stack) turn}\footnote{In some literature, the term ``turn'' has been also used to indicate a change of the direction of an input/work tape head move. In this work, we do not discuss such a turn.}
of a pushdown automaton $M$ is a series of actions of $M$ during the time when the mode of $M$'s  stack height is increasing and then decreasing. A \emph{$k$-turn pushdown automaton} is a pushdown automaton that makes at most $k$ turns on each computation path on every input.

\subsection{Nonuniform Families of Machines}\label{sec:FL-FLpoly}

In what follows, we consider a family $\LL=\{(L_n^{(+)},L_n^{(-)})\}_{n\in\nat}$  of promise (decision) problems over an alphabet, say, $\Sigma$. To solve such a promise problem family, we wish to focus on a family $\MM=\{M_n\}_{n\in\nat}$ of machines of the same type (that is, 1dfa's, 1nfa's, etc.).
For those two families $\LL$ and $\MM$, we say that $\MM$ \emph{solves} $\LL$ if, for each index $n\in\nat$, $M_n$ solves $(L_n^{(+)},L_n^{(-)})$.
As remarked in Section \ref{sec:introduction}, we particularly concentrate on the ``size'' and the ``ceiling'' of automata families throughout this work.

A family $\MM=\{M_n\}_{n\in\nat}$ of finite automata (resp., pushdown automata) $M_n$ with sets $Q_n$ of inner states and stack alphabets $\Gamma_n$ with push sizes $e_n$ is said to have \emph{polynomial size} if there exists a polynomial $p$ satisfying $|Q_n|\leq p(n)$ (resp., $|Q_n||\Gamma_n^{\leq e_n}|\leq p(n)$) for all $n\in\nat$.


The notation $\oned$ is used for the collection of all families of promise problems solvable by families of polynomial-size 1dfa's. Similarly, the notation $\onen$ is used for its nondeterministic variant. We also define $\onebp$ and $\onep$ by demanding all underlying 1pfa's to make \emph{bounded-error probability} and \emph{unbounded-error probability}, respectively. For the case of \emph{one-side-error probability}, we write $\oner$.
As for the use of pushdown automata families, we define $\onedpd$ (resp., $\onenpd$) to be the collection of all families of promise problems solvable by families of polynomial-size 1dpda's (resp., 1npda's). In a similar fashion, we define $\twod$, $\twon$, $\twodpd$, and $\twonpd$ using 2dfa's, 2nfa's, 2dpda's, and 2npda's, respectively.
When the behaviors of 1dpda's are limited to making only at most $k$ turns, we use the notation of $k\mathrm{t1DPD}$ instead of $\onedpd$.
For a family $\LL=\{(L_n^{(+)},L_n^{(+)})\}_{n\in\nat}$, its \emph{complement}, denoted $\co\LL$, is $\{(L_n^{(-)},L^{(+)})\}_{n\in\nat}$. Given a class $\CC$ of promise problem families, the notation $\co\CC$ denotes the set $\{\LL\mid \co\LL\in\CC\}$.

Now, we wonder how finite and pushdown automata families can be empowered by their machines'  sizes. Beyond the polynomial-size restriction,
we also target superpolynomial-size automata families.
Earlier, Kapoutsis \cite{Kap09,Kap12} invented the notations of $2^{\twod}$ and $2^{\twon}$ to treat the case of  ``exponential-size''.
More generally, given a univariate ``function'' $f$, we introduce the generic notation of $2^{f(\twod)}$ (resp., $2^{f(\twon)}$)
by simply taking families of 2dfa's (resp., 2nfa's) of $2^{O(f(n^{O(1)}))}$  size.
For example, if we take $f(n)=n$ and $f(n)=\log^k{n}$ for $2^{f(\twod)}$, then we obtain the complexity classes $2^{\twod}$ and $2^{\log^k{\twod}}$, respectively.
As for pushdown automata families, we also introduce the notation $2^{f(\twodpd)}$ (resp., $2^{f(\twonpd)}$) using families of $2^{O(f(n))}$-size  2dpda's (resp., 2npda's). In a similar manner, we also define $2^{\linoner}$ using one-sided-error 1pfa's of linear-exponential (i.e., $2^{O(n)}$) size.

A \emph{ceiling} is an important concept, which refers to a length bound of input strings given to each promise problem. A family $\LL$ over $\Sigma$ is said to have \emph{polynomial ceiling} (resp., \emph{exponential ceiling}) if there exists a polynomial $p$ such that,  for all numbers $n\in\nat$, $p(n)$ (resp., $2^{p(n)}$) upper-bounds the length of valid inputs, namely,  $L_n^{(+)}\cup L_n^{(-)}\subseteq \Sigma^{\leq p(n)}$ (resp., $L_n^{(+)}\cup L_n^{(-)}\subseteq \Sigma^{\leq 2^{p(n)}}$).
Kapoutsis \cite{Kap09,Kap12} again invented the notation $\twod/\poly$ (resp., $\twod/\mathrm{exp}$) for the restriction of $\twod$ onto families of promise problems having polynomial (resp., exponential) ceilings. Similarly, we can introduce the notations, such as $\onedpd/\poly$ and $\twonpd/\mathrm{exp}$.
For the sake of later convenience,  $\mathrm{ALL}/\poly$ (resp., $\mathrm{ALL}/\mathrm{exp}$) expresses the collection of all families of promise problems with polynomial (resp., exponential) ceilings. In the rest of this work, our targets are subclasses of these collections.

Whenever we need to restrict underlying machines to run in \emph{polynomial time} (resp., \emph{exponential time}) in both $n$ and $|x|$, we use the special prefix of ``ptime-'' (resp., ``etime-'') as in, e.g., $\mathrm{ptime\mbox{-}}2^{\twod}$ (resp., $\etime2^{\log^k{\twodpd}}$).
It is important to remark that, as shown in \cite{GMP03}, underlying 2dfa's and 2nfa's can be all treated to have  ``polynomial'' runtime.
By sharp contrast, various runtime restrictions of 2dpda's and 2npda's are crucial for the computational complexity of solvable promise problems.

The (partial) function class $\mathrm{1F}$ was introduced in \cite{Yam23}. This work further introduces its two-way version, which we call $\twof$.
A family $\{(f_n,D_n)\}_{n\in\nat}$ of partial functions from $\Sigma^*$ to $\Gamma^*$ for two alphabets $\Sigma$ and $\Gamma$ is in $\twof$ if there exist a polynomial $p$ and a family $\{M_n\}_{n\in\nat}$ of 2dft's such that, for any index $n\in\nat$ and any string $x\in D_n$, $M_n$ takes $\rhd{x}\lhd$ on its input tape and produces $\rhd{f_n(x)}$ on its write-once output tape in time $p(n,|x|)$ until $M_n$ finally enters a halting (inner) state.
As before, we call those 2dft's by ``polynomial-time'' 2dft's.
This runtime bound naturally  forces the length of output strings to be polynomially bounded.

\subsection{Counter Automata and Counter Pushdown Automata}\label{sec:counter}

In the setting of nonuniform (stack-)state complexity, the power of multiple counters was discussed in \cite{Yam24,Yam24b}.
When a stack uses its stack alphabet $\Gamma$ consisting only of a single non-bottom symbol, say, ``$1$'' (i.e., $\Gamma=\{1,\bot\}$), it is specifically called a  \emph{counter}. A \emph{$k$-counter automaton} (resp., a \emph{$k$-counter pushdown automaton}) is a finite automaton (resp., a pushdown automaton) equipped with $k$ counters. See, e.g., \cite{Yam23b,Yam24,Yam24b} for their basic behaviors.
When underlying machines use $k$ counters, we use the notations of $\onedct_k$, $\onenct_k$, $\onedpdct_k$, and $\onenpdct_k$, where the suffix ``$\mathrm{CT}_k$'' refers to the use of $k$ counters.
In the case of two-way head moves, we replace the prefix ``$1$'' by ``$2$'', such as $\twodct_k$, $\ptime\twofct_k$, $\twodpdct_k$, etc. In particular, when $k$ equals $1$, we tend to drop the subscript ``$k$''.
In this work, we further provide deterministic finite transducers with multiple counters and introduce the complexity classes $\twofct_k$ and $\ptime\twofct_k$.

\subsection{Random Access to Karp-Lipton Style Advice}

Throughout this work, we assume the reader's familiarity with deterministic Turing machine (DTM) and its nondeterministic variant (NTM) as well as Cook's \cite{Coo71} model of \emph{auxiliary pushdown automata}, which  are two-way pushdown automata equipped with rewritable auxiliary (work) tapes used as an extra memory device. For simplicity, we abbreviate  a nondeterministic (resp., deterministic) auxiliary pushdown automaton as an \emph{naux-pda} (resp., a \emph{daux-pda}).

Let us recall that
$\mathrm{SC}^k$ denotes the $k$th Steve's Class and
its nondeterministic variant $\mathrm{NSC}^k$.
We introduce a more general notation of $\nauxpdaspti{s(n)}{t(n)}$ (resp., $\dauxpdaspti{s(n)}{t(n)}$) for the complexity class induced by the use of naux-pda's (resp., daux-pda's) running in time $t(n)$ using space $s(n)$.
With this notation, $\mathrm{SC}^k$ (resp., $\mathrm{NSC}^k$) coincides with $\dauxpdaspti{O(\log^kn)}{n^{O(1)}}$ (resp., $\nauxpdaspti{O(\log^kn)}{n^{O(1)}}$).

We further equip Karp-Lipton style advice to daux-pda's and naux-pda's. We the  use of $h(n)$-size advice strings, we expand $\mathrm{SC}^k$ (resp., $\mathrm{NSC}^k$) to $\mathrm{SC}^k/h(n)$ (resp., $\mathrm{NSC}^k/h(n)$).
Here,  the suffix ``$/h(n)$'' refers to the use of Karp-Lipton style advice of size $O(h(n))$.

In particular, we pay special attention to the case of $h(n)=2^{O(\log^k{n})}$.
It is important to remark that, because the advice has size $2^{O(\log^kn)}$ for inputs of length $n$, the polynomial  runtime restriction
hinders  underlying Turing machines  from reading all
symbols of each advice string, and thus we use a standard convention of making a \emph{random access}\footnote{Whenever a machine intends to access the $i$th tape cell, it first produces the binary number expressing ``$i$'' on an index tape and then enters a designated ``query'' state so that the content of the $i$th tape cell is automatically retrieved and written directly on an ``answer'' tape.}
(instead of moving a tape head sequentially back and forth) whenever accessing such a long advice string.

\section{Two Key Supporting Technical Tools}\label{sec:tools}

For later expositions, at this point, we wish to explore two key supporting technical tools necessary for proving various containment results of nonuniform complexity classes, as shown in Fig.~\ref{fig:hierarchy-poly} and \ref{fig:hierarchy-exp}.

\subsection{Reductions by Two-Way Deterministic Finite Transducers}\label{sec:reductions}

In computational complexity theory, reductions have played a significant role in comparing the computational complexity of two separate computational problems. With a similar spirit, we
wish to utilize the notion of ``reductions'' between two families of promise problems.
Earlier, Sakoda and Sipser \cite{SS78} discussed ``homomorphic reducibility,'' which was later used by Kapoutsis \cite{Kap09,Kap12,Kap14} as a reduction tool to identify the most difficult promise problem families.
Here, we introduce a notion of more powerful reductions.
We stress that the reduction notion is interesting in its own right in the study of nonuniform automata families and it surely requires further intensive research to prove its usefulness.

Let $\LL=\{(L_n^{(+)},L_n^{(-)})\}_{n\in\nat}$ and $\KK=\{(K_n^{(+)},K_n^{(-)})\}_{n\in\nat}$ denote two families of promise problems over alphabets $\Sigma$ and $\Gamma$, respectively. We say that $\LL$ is \emph{ptime-2FCT$_k$ many-one reducible} (or \emph{ptime-2FCT$_{k}$-m-reducible}, for short) to $\KK$ (denoted $\LL\leq^{\ptime\twofct_k}_{m}\KK$) if there exist a family $\{(f_n,D_n)\}_{n\in\nat}$ of partial functions in $\ptime\twofct_k$ and another p-bounded, p-honest function $d:\nat\to\nat$ that satisfy the following three conditions for any $n\in\nat$ and for any $x\in\Sigma^*$:  (i) $L_n^{(+)}\cup L_n^{(-)}\subseteq D_n$, (ii) $x\in L_n^{(+)}$ implies $f_n(x)\in K_{d(n)}^{(+)}$, and (iii) $x\in L_n^{(-)}$ implies $f_n(x)\in K_{d(n)}^{(-)}$.
When neither counters nor time bounds are required for reduction machines, we instead write $\LL\leq^{\twof}_{m}\KK$.

Given a family $\LL$ of promise problems, the special notation $\leq^{\ptime\twofct_k}_{m}\!(\LL)$ denotes the collection of all families of promise problems that are 2FCT$_{k}$-m-reducible to $\LL$.
Moreover, for a class $\CC$ of families of promise problems, $\leq^{\ptime\twofct_k}_{m}\!(\CC)$ expresses the union $\bigcup_{\LL\in\CC}\leq^{\ptime\twofct_k}_{m}\!(\LL)$. We also define $\leq^{\twof}_{m}\!(\LL)$ if neither counters nor time bounds are needed.

It is known in
\cite{Yam24,Yam24b} that
(1) for any $k\geq4$, $\twodct_k=\twodct_4$,
(2) for any $k\geq3$, $\twodpdct_k=\twodpdct_3$, and
(3) $\twodct_4/\poly = \twod/\poly$ and $\twodpdct_3/\poly = \twodpd/\poly$. The nondeterministic cases also hold for (1)--(3).
Moreover, it is shown in \cite{Yam24,Yam24b} that
$\twod\neq \twodct$, $\twou\neq \twouct$, and $\twon\neq \twonct$, where ``U'' indicates ``unambiguous'' computation.

\begin{proposition}\label{reduction-closure}
(1) $\ptime\twodpdct_{3}  = \: \leq^{\ptime\twofct_2}_{m}\!(\ptime\twodpdct_{3})
= \: \leq^{\ptime\twofct_2}_{m}\!(\onedpdct_3)$.
(2) $\ptime\twodct_{4} = \: \leq^{\ptime\twofct_2}_{m}\!(\ptime\twodct_{4})
=  \: \leq^{\ptime\twofct_2}_{m}\!(\onedct_4)$.
The same equalities hold for nondeterministic computations.
\end{proposition}

We remark that the inclusions $\ptime\twodpdct_3\subseteq \: \leq^{\ptime\twofct_2}_{m}\!(\onedpdct_3)$ and $\ptime\twonpdct_3 \subseteq \: \leq^{\ptime\twofct_2}_{m}\!(\onenpdct_3)$ are reminiscent of \cite[Lemma 3]{Sud78}, in which the language recognized by a polynomial-time two-way pushdown automaton can be $\dl$-m-reducible to an appropriate context-free language.

In Proposition  \ref{reduction-closure}(2), when inputs provided to underlying machines are polynomially bounded, we can eliminate the full use of ``counters'', even from reduction functions.

\begin{corollary}\label{twod-poly-twodpd}
(1) $\twod/\poly  = \: \leq^{\twof}_{m}\!(\oned/\poly)
= \: \leq^{\twof}_{m}\!(\twod/\poly)$.
(2) $\ptime\twodpd/\poly  = \: \leq^{\twof}_{m}\!(\onedpd/\poly)
= \: \leq^{\twof}_{m}\!(\ptime\twodpd/\poly)$.
\end{corollary}

By an instant application of Corollary \ref{twod-poly-twodpd}, we conclude that, if $\oned/\poly=\onen/\poly$, then $\twod/\poly=\twon/\poly$, which is unknown to hold without any assumptions.
See also Proposition \ref{one-way-case-collapse}(1).

Beyond the above corollary, it is not clear that
$\ptime\twodpd =  \: \leq^{\twof}_{m}\!(\ptime\twodpd)$
and $\twod =  \; \leq^{\twof}_{m}\!(\twod)$.

In a similar vein to Proposition \ref{reduction-closure}, we further assert the following closure property.

\begin{proposition}\label{closure-log-k}
$\ptime2^{\log^k\twodct_{4}}   = \: \leq^{\ptime\twofct_2}_{m}\!(\ptime2^{\log^k\twodct_{4}}) = \: \leq^{\ptime\twofct_2}_{m}\!(2^{\log^k\onedct_{4}})$.
\end{proposition}

Similarly to Corollary \ref{twod-poly-twodpd}, we obtain from Proposition \ref{closure-log-k} the following result, in which the use of counters and the runtime bounds in the proposition are eliminated.

\begin{corollary}\label{log-k-2D-poly}
$2^{\log^k\twod}/\poly
=\: \leq^{\twof}_{m}\!(2^{\log^k\twod}/\poly) =\: \leq^{\twof}_{m}\!(2^{\log^k\oned}/\poly)$.
\end{corollary}

\subsection{Translation between Automata Families and Advised Turing Machines}

The second technical tool is to utilize underlying connections between families of finite automata and advised Turing machines and also between pushdown automata families and advised auxiliary pushdown automata.
Such connections were first discussed by Berman and Lingas \cite{BL77} and by Sakoda and Sipser \cite{SS78} and have been in a central subject of recent research.

These connections help us exploit collapse/separations of (mainstream) complexity classes.
Here, we present two technical propositions, which will be frequently used in later sections. We say that a function on $\nat$  is \emph{2dft-constructible} if there is a 2dft $M$ such that $M$ on each input of the form $1^n$ produces the string $1^{f(n)}$ on its write-once output tape.
The first key proposition is given as follows.

\begin{proposition}\label{key-proposition}
Let $f,g,s,d$ be any functions on $\nat$   with $f(n)\geq1$ for all $n\in\nat$.  Assume that $d$ is nondecreasing and that $f$ is 2dft-constructible. Assume also that the inverse $f^{-1}$ of $f$ exists on the image of $f$. Let $\LL$ denote a family $\{(L_n^{(+)},L_n^{(-)})\}_{n\in\nat}$ of promise problems over an alphabet $\Sigma$. Assume that $\LL$ has $f(n)$-ceiling.

(1) If $\LL$ is solved by a family $\{N_n\}_{n\in\nat}$ of $g(n)$-size 2-choice 2nfa's (resp., 2npda's), then there exists an NTM (resp., an naux-pda) $N$ such that $N$ recognizes the language $K$ using $O(\log{e(|w|)})$ space with the help of $O(e(|w|))$-size advice, where $K  =\{1^{f(n)-|x|-1}\# x\mid n\in\nat, |x|\leq f(n), x\in L(N_n)\}$ and $e(|w|) = g\circ f^{-1}(|w|)$.

(2) If the above-defined language $K$ is recognized by an NTM (resp., an naux-pda) $M$ using $s(|w|)$ space with $d(|w|)$-size advice for inputs $w$, then
(i) for each $n$, $L_n^{(+)} \subseteq \{x\mid |x|\leq f(n), 1^{f(n)-|x|-1}\# x\in K\}$ and $L_n^{(-)} \subseteq \{x\mid |x|\leq f(n), 1^{f(n)-|x|-1}\# x\in \overline{K}\}$ and (ii) there exists a family $\{M_n\}_{n\in\nat}$ of $e'(n)$-size 2nfa's (resp., 2npda's) solving $\LL$, where $e'(n) = 2^{O(s\circ f(n))}+O(d\circ f(n))$.

(3) The above statements (1)--(2) also hold in the deterministic case.
\end{proposition}

The following proof argument of Proposition \ref{key-proposition} looks quite different from the arguments given in \cite{Yam19a,Yam21,Yam24,Yam24b}, where \cite{Yam24b} particularly provides a standard, general framework to those arguments.


The first key proposition (Proposition \ref{key-proposition}) intends to first translate finite automata (as well as pushdown automata) into NTMs (as well as naux-pda's). Conversely, the following second key proposition first translates NTMs (and naux-pda's) to finite automata (and pushdown automata),  compensating the first key proposition.

\begin{proposition}\label{second-key-proposition}
Let $f,g,s,d$ be any functions on $\nat$  with $f(n)\geq1$ for all $n\in\nat$. Assume that $d$ is nondecreasing and that $f$ is 2dft-constructible. Assume also that the inverse $f^{-1}$ of $f$ exists on the image of $f$. Let $K$ denote any language over an alphabet $\Sigma$.

(1) If $K$ is recognized by an NTM (resp., an naux-pda) $N$ using $s(|x|)$ space with $d(|x|)$-size advice for inputs $x$, then there exists a family $\{N_n\}_{n\in\nat}$ of $e(n)$-size $O(1)$-choice 2nfa's (resp., 2npda's)  solving $\LL=\{(L_n^{(+)},L_n^{(-)})\}_{n\in\nat}$, where  $L_n^{(+)}=\{ x\mid |x|=f(n), x\in K\}$,  $L_n^{(-)}=\{x\mid |x|= f(n), x\in \overline{K}\}$, and $e(n)=2^{O(s\circ f(n))}+O(d\circ f(n))$. This family $\LL$ has $f(n)$-ceiling.

(2) If the above-defined family $\LL$ is solved by a family $\{M_n\}_{n\in\nat}$ of $g(n)$-size 2nfa's (resp., 2npda's), then (i) $K$ is identified with $\{x\mid n\in\nat, |x|=f(n), x\in L(M_n)\}$ and (ii) there exists an NTM (resp., an naux-pda) $M$ that recognizes the language $K$ using $O(\log{e'(|x|)})$ space with the help of $O(e'(|x|)^2)$-size advice for inputs $x$, where $e'(|x|)= g\circ f^{-1}(|x|)$.

(3) The statements (1)--(2) also hold in the deterministic case.
\end{proposition}

\section{Promise Problem Families of Polynomial Ceilings}\label{sec:poly-ceilings}

\begin{figure}[t]
\centering
\resizebox*{!}{5.2cm}{\includegraphics*{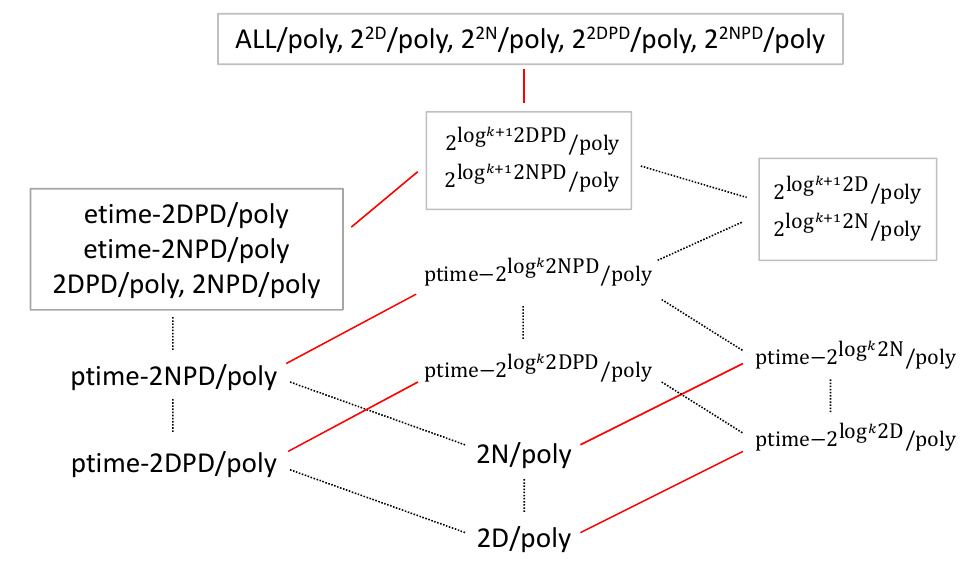}}
\caption{Inclusion relationships among nonuniform complexity classes with polynomial ceilings except for the separation result of Theorem \ref{ptime-log-k-subset}. Solid red lines indicate proper inclusions. The superscript ``$k$'' refers to an arbitrary integer at least $2$.}\label{fig:hierarchy-poly}
\end{figure}

Let us first recall the purpose of this work, which is to study the effect of various ceilings (as well as various sizes) on the overall complexity of families of promise problems.
Throughout this section, we wish to study the computational complexity of nonuniform automata families particularly having polynomial ceilings. Most of our results are illustrated in Fig.~\ref{fig:hierarchy-poly}.

Kapoutsis \cite{Kap14} is the first to focus on the polynomial ceilings when he demonstrated, assuming the ``strong advice'' restriction, that $\twod/\poly = \twon/\poly$ exactly when  $\nl\subseteq \dl/\poly$.
This obviously hints that $\twod/\poly$ and $\twon/\poly$ may differ. Under no unproven assumption, by clear contrast,  it follows that $\twou/\poly = \twon/\poly$ and $\ptime\twoupd/\poly = \ptime\twonpd/\poly$   \cite{Yam24,Yam24b}, where ``$\mathrm{U}$'' indicates the ``unambiguous'' computations of underlying finite and pushdown automata.

\subsection{Complexity of Polynomial-Size Automata Families}\label{sec:poly-size-automata}

Sudborough \cite{Sud78} demonstrated that all languages in $\logcfl$ can be simulated by naux-pda's in polynomial time using $O(\log{n})$ space.
Ruzzo \cite{Ruz80} proved that such a language can be further accepted by an appropriate daux-pda running in $O(\log{n})$ space and $2^{O(\log^2n)}$ time. 
This instantly implies that $\ptime\twonpd \subseteq \etime\twodpd$ with no ceiling bound.
From a different perspective, we wish to discuss upper bounds of $\twodpd$ and $\twonpd$ in terms of $\twod$ and $\twon$ when all promise problem families are restricted to polynomial ceiling.

\begin{theorem}\label{Cook-proof}
(1) $\twod/\poly \subseteq \ptime\twodpd/\poly \subseteq \ptime2^{\log^2\twod}/\poly$.

(2) $\twon/\poly \subseteq \ptime\twonpd/\poly \subseteq \ptime2^{\log^2\twon}/\poly$.
\end{theorem}

With no ceiling bound, on the contrary, it is still unknown whether $\ptime\twodpd\subseteq 2^{\log^2\twod}$ holds or even $\ptime\twodpd\subseteq 2^{\lintwod}$ holds.

Let us return to the proof of Theorem \ref{Cook-proof} and give its proof with the help of Corollaries  \ref{twod-poly-twodpd} and \ref{log-k-2D-poly}.

\vs{-2}
\begin{proofsketchof}{Theorem \ref{Cook-proof}}
(1) By Corollaries \ref{twod-poly-twodpd} and \ref{log-k-2D-poly}, we obtain $\ptime\twodpd/\poly =  \; \leq^{\twof}_{m}\!(\onedpd/\poly)$ and $2^{\log^k\twod}/\poly = \:  \leq^{\twof}_{m}\!(2^{\log^k\oned}/\poly)$. Thus, it suffices to prove that $\onedpd/\poly \subseteq 2^{\log^2\twod}$.
An underlying idea of this proof comes from Cook's proof that all deterministic context-free languages are in $\mathrm{SC}^2$
\cite{Coo71} (see also \cite{BCM+83}).

(2) Similarly to (1), it suffices to prove that $\onenpd/\poly \subseteq 2^{\log^2\twon}$. Earlier, Ruzzo \cite[Corolalry 7]{Ruz80} argued that the maximal stack height on each computation path of a 1npda can be reduced to $O(\log^2n)$, where $n$ is input length.
A similar proof shows that a polynomial-time 2npda $M_n$ having $n$ inner states can be simulated by another polynomial-time 2npda having $n^{O(1)}$ inner states using only $O(\log^2|x|)$ stack height. We then encode all possible stack contents of height $O(\log^2n)$  into inner states.
\end{proofsketchof}


By Theorem \ref{Cook-proof}, $\twon/\poly \subseteq \ptime\twonpd/\poly$ holds; however, it is not yet known whether the opposite inclusion holds.  This is in fact boiled down to the $\onenpd/\poly \subseteq?\twon$ question, as shown below.

\begin{proposition}
$\twon/\poly = \ptime\twonpd/\poly$ if and only if $\onenpd/\poly \subseteq \twon/\poly$.
\end{proposition}

\begin{yproof}
If $\onenpd/\poly \subseteq \twon/\poly$, then $\leq^{\twof}_{m}\!(\onedpd/\poly) \subseteq  \: \leq^{\twof}_{m}\!(\twon/\poly)$, which implies that $\ptime\twonpd/\poly \subseteq \twon/\poly$ by Corollary \ref{twod-poly-twodpd}. Since $\twon/\poly \subseteq \ptime\twonpd/\poly$ is obvious, it thus follows that $\twon/\poly = \ptime\twonpd/\poly$.
\end{yproof}


We have already discussed the complexity class $\ptime\twonpd/\poly$.  It was shown in \cite{Yam21} that $\ptime\twodpd/\poly = \ptime\twonpd/\poly$ if and only if $\logdcfl/\poly = \logcfl/\poly$, where $\logcfl$ denotes the closure of $\cfl$ (context-free language class) under polynomial-time many-one reductions and $\logdcfl$ is its deterministic variant. However,  this situation may differ in the case of no ceiling bound.

\begin{proposition}\label{twonpd-with-poly}
$\twonpd/\poly = \twodpd/\poly = \etime\twodpd/\poly = \etime\twonpd/\poly$.
\end{proposition}

 \begin{proofsketch}
It is obvious that  $\etime\twodpd/\poly \subseteq \etime\twonpd/\poly \subseteq \twonpd/\poly$, and $\etime\twodpd/\poly \subseteq \twodpd/\poly \subseteq \twonpd/\poly$.
It thus suffices to verify that $\twonpd/\poly \subseteq \etime\twodpd/\poly$.
For any family $\LL=\{(L_n^{(+)},L_n^{(-)})\}_{n\in\nat}$ having a polynomial ceiling over an alphabet $\Sigma$, take a family $\{N_n\}_{n\in\nat}$ of polynomial-size 2npda's solving $\LL$.
By Proposition \ref{key-proposition}(1) with a suitable polynomial $f$, the language $K=\{1^{f(n)-|x|-1}\#x\mid n\in\nat, |x|\leq f(n), x\in L(M_n)\}$ belongs to $\nauxpdaspti{O(\log{n})}{\infty}/\poly$.  By an advised version of Cook's result \cite{Coo71}, $K$ also belongs to $\dauxpdaspti{O(\log{n})}{\infty}/\poly$.
Ruzzo \cite{Ruz80} claimed that the time complexity of underlying daux-pda's can be reduced to exponentials (i.e., $2^{n^{O(1)}}$) without changing this complexity class.  By Proposition \ref{key-proposition}(2),  $\LL$ falls in $\etime\twodpd/\poly$.
\end{proofsketch}

We remark that it is not known whether $\twodpd = \twonpd$ holds.

\subsection{Complexity of Superpolynomial-Size Automata Families}\label{sec:exp-size-automata}

In contrast to Section \ref{sec:poly-size-automata},  we wish to explore the roles of various nonuniform automata families of superpolynomial size.

We start with the following easy claim.

\begin{proposition}\label{ALL-equivalent}
$\mathrm{ALL}/\poly = 2^{\twod}/\poly = 2^{\twon}/\poly = 2^{\twodpd}/\poly = 2^{\twonpd}/\poly$.
\end{proposition}

\begin{yproof}
It suffices to prove that $\mathrm{ALL}/\poly \subseteq 2^{\twod}/\poly$. We take any family $\LL=\{(L_n^{(+)},L_n^{(-)})\}_{n\in\nat}$ in $\mathrm{ALL}/\poly$ over an alphabet $\Sigma$. There is a polynomial $p$ satisfying  $L_n^{(+)}\cup L_n^{(-)}\subseteq \Sigma^{\leq p(n)}$ for all $n\in\nat$.
For each index $n\in\nat$, we define a 2dfa $M_n$ as follows. The set $Q_n$ consists of all tuples of the from $(j,s_1s_2 \cdots s_j,b)$ with $j\in[p(n)]$, $s_1,\ldots,s_j\in\Sigma$, and $b=L_n^{(+)}(s_1s_2\cdots s_j)$, where $L_n^{(+)}(x)$ denotes the acceptance (1) or the rejection (0) of $x$ to $L_n^{(+)}$.
Consider the following algorithm.
On input $x$ of the form $s_1s_2\cdots s_m$ with $s_i\in\Sigma$, we process input symbols $s_1,s_2,\ldots,s_m$ one by one together with changing inner states $(1,s_1,b_1),(2,s_1s_2,b_2),\ldots,(m,s_1s_2\cdots s_m,b_m)$ in this order, and finally output $b_m$. This algorithm can be implemented on a suitable 2dfa with $Q_n$ as a set of inner states.
Since  $|Q_n|\leq 2^{O(p(n))}$, we conclude that  $\LL$ belongs to $2^{\twod}/\poly$.
\end{yproof}


Theorem \ref{Cook-proof} is extendable to $2^{O(\log^k{n})}$-size automata families for various values of $k\geq2$.

\begin{theorem}\label{log-hierarchy-relation}
For any integer $k\geq2$, $\ptime2^{\log^k\twon}/\poly \subseteq \ptime2^{\log^k\twonpd}/\poly \subseteq 2^{\log^{k+1}\twod}/\poly$.
\end{theorem}

\begin{yproof}
The first inclusion is obvious. Thus, we focus on the second inclusion between  $\ptime2^{\log^k\twonpd}/\poly$ and $2^{\log^{k+1}\twod}/\poly$.

Take an arbitrary family $\LL=\{(L_n^{(+)},L_n^{(-)})\}_{n\in\nat}$ in $\ptime2^{\log^k\twonpd}/\poly$ over alphabet $\Sigma$. There are two constants $a,c>0$ and a family $\{M_n\}_{nb\in\nat}$ of $2^{c\log^kn+c}$-size 2npda's solving $\LL$ in time $(n|x|)^c+c$. We also assume that $\LL$ is of $(n^a+a)$-ceiling.  We set $f(n)=n^a+a$ and $g(n)=2^{c\log^kn+c}$. We then define the language $K$ as $\{1^{f(n)-|x|-1}\# x\mid n\in\nat,|x|\leq f(n),x\in L(M_n)\}$. Without loss of generality, we assume by Lemma \ref{two-choice-change} that $M_n$ makes at most $2$-choices. Note that $e(n)=g\circ f^{-1}(n) = 2^{c\log^k(n-a)^{1/a}+c}\leq 2^{c'\log^kn+c'}$ for a suitable constant $c'>0$.
Proposition \ref{key-proposition}(1) provides an naux-pda $N$ for $K$. ,
Since the runtime of $M_n$ on input $x$ is $(n|x|)^c+c$ for a constant $c>0$, $N$ runs in time $n^{O(1)}$. Since $\log{e(|w|)}=O(\log^k|w|)$, $N$ uses space $O(\log^k|w|)$.
Therefore, $K$ falls in $\nauxpdaspti{O(\log^kn)}{n^{O(1)}}/2^{O(\log^kn)}$.

As Ruzzo \cite{Ruz80} demonstrated, $\nauxpdaspti{O(\log^kn)}{n^{O(1)}}$ is included in $\mathrm{DSPACE}(O(\log^{k+1}n))$. We remark that his proof can be carried out even in the presence of advice.  We thus immediately obtain $K\in \mathrm{DSPACE}(O(\log^{k+1}n))/2^{O(\log^kn)}$. There exists  a DTM $M$ that recognizes $K$ using space at most $c\log^{k+1}n+c$ with $2^{a\log^kn+a}$-size advice for suitable constants $a,c>0$.
We set $s(n)=c\log^{k+1}n+c$ and $d(n)=2^{a\log^kn+a}$. Note that $e'(n)=2^{O(\log^{k+1}n)}+2^{O(\log^kn)}=2^{O(\log^{k+1}n)}$ because $s\circ f(n)=O(\log^{k+1}n)$, and $d\circ f(n)=2^{\log^kn)}$. By Proposition \ref{key-proposition}(2), $\LL$ belongs to $2^{\log^{k+1}\twod}/\poly$.  \end{yproof}

Theorem \ref{log-hierarchy-relation} indicates that the complexity of $\ptime2^{\log^k\twon}/\poly$ is relatively high. How high is it? Is it higher than, say, $\twodpd/\poly$? As a partial answer to this question, we demonstrate that  at least $\ptime2^{\log^k\twod}/\poly$ is not ``small'' enough to be included in $\twonpd/\poly$.

\begin{theorem}\label{ptime-log-k-subset}
$\ptime2^{\log^2\twod}/\poly \nsubseteq \twonpd/\poly$.
\end{theorem}

\begin{proofsketch}
Toward the verification of the theorem, we wish to  claim that (1)  $\ptime2^{\log^2\twod}/\poly \subseteq \twonpd/\poly$ implies $\mathrm{SC}^2/2^{O(\log^2{n})} \subseteq \p/\poly$ and (2) $\mathrm{SC}^2/2^{O(\log^2{n})}\nsubseteq \p/\poly$.

(1) This is proven in a way similar to the proof of Theorem \ref{log-hierarchy-relation}.

(2) This argument proceeds by a diagonalization technique.
We first enumerate all DTMs as  $M_1,M_2,\ldots$, all binary strings $x_{1,m},x_{2,m},\ldots$, and all binary advice strings $a_{1,m},a_{2,m},\ldots$ with $m\in\nat^{+}$,
where $|x_{j,m}|=|a_{j,m}|=m$ with $1\leq j\leq 2^m$.
It is possible, without loss of generality, that we fixate the advice alphabet to be $\{0,1\}$.
Given $(i,j,l)$, we write the outcome (0 or 1) of $M_i$ on input $x$ together with advice string $a_{j,|x|^l}$ within time $|x|^l$ as $M_i(x; a_{j,|x|^l})$ if it exists; otherwise, we automatically set $M_i(x;a_{j,|x|^l})$ to be $\lambda$, where we identify
1 and 0 with ``acceptance'' and ``rejection'', respectively.

We define an advice function $h(m)$ and a language $L$ as follows.
We set $h(m)= b_1b_2\cdots b_{m^l} \# 1^i\# 1^l \# z$ if $z\in\{0,1\}^*$, $m=\pair{i,l}01^t$
for $i,l\leq \log{m}$, $t\in\nat$, and $b_j=M_i(x_{j,m},a_{j,m^l})$ with $1\leq j\leq m^l$ and $i+l+m^l+|z|+3=2m^l$,  where $\pair{\cdot}$ denotes an appropriate coding function.
In particular, it follows that the $j$th symbol of $h(m)$ equals $1$ iff $M_i$ accepts $x_{j,m}$  using $a_{j,m^l}$ within $m^l$ steps.

Now, we define the example language $L$ to be the set $\{x_{j,m} \mid \text{ the $j$th symbol of $h(|x_{j,m}|)$ is not $1$ }\}$. We then verify that $L\notin \p/\poly$ by leading to a contradiction from $L\in\p/\poly$.
\end{proofsketch}

From Theorem \ref{ptime-log-k-subset}, we immediately obtain the following two corollaries.

\begin{corollary}
(1) $\ptime2^{\log^2\twon}/\poly \nsubseteq \twon/\poly$.
(2) $\ptime2^{\log^2\twod}/\poly \nsubseteq \twod/\poly$.
\end{corollary}

\begin{corollary}
(1) $\ptime2^{\log^2\twonpd}/\poly \nsubseteq \ptime\twonpd/\poly$.
(2) $\ptime2^{\log^2\twodpd}/\poly \nsubseteq \ptime\twodpd/\poly$.
\end{corollary}


Since $\ptime2^{\log^k\twon}/\poly$ is not ``small'' by Theorem \ref{ptime-log-k-subset}, we further wonder if  $\ptime2^{\log^k\twod}/\poly$ and $\ptime2^{\log^k\twon}/\poly$ coincide.
It then turns out by the following proposition that they may not coincide because the collapse of them is related to answering the $\mathrm{NSC}^{k}\subseteq$?$\mathrm{SC}^{k}/2^{O(\log^kn)}$ question, which is expected to fail.

\begin{proposition}\label{log-k-twod-SC}
For each integer with $k\geq2$,
$\ptime2^{\log^k\twod}/\poly = \ptime2^{\log^k\twon}/\poly$ if and only if $\mathrm{NSC}^k \subseteq \mathrm{SC}^k/2^{O(\log^kn)}$.
\end{proposition}

\begin{proofsketch}
(If--part) Assume that $\mathrm{NSC}^k\subseteq \mathrm{SC}^k/2^{O(\log^kn)}$.
We remark that this inclusion relation is logically equivalent to $\mathrm{SC}^k/2^{O(\log^k{n})} = \mathrm{NSC}^k/2^{O(\log^k{n})}$.
Now, we begin with an arbitrary family $\LL=\{(L_n^{(+)},L_n^{(-)})\}_{n\in\nat}$ in $\ptime2^{\log^k\twon}/\poly$ over an alphabet $\Sigma$.
Take a polynomial $f$ for which $\LL$ has $f(n)$-ceiling.
There exists a family $\{N_n\}_{n\in\nat}$ of $2^{O(\log^k{n})}$-size 2dfa's solving $\LL$ in time polynomial in $(n,|x|)$.
Assume that $N_n$ takes an input $x$ using at most $2^{a \log^kn+a}$ inner states and runs in $n^a+a$ time for a suitable constant $a\in\nat^{+}$.  For simplicity, we set $g(n)= 2^{a \log^k{n}+a}$ for all $n\in\nat$ and assume that $f(n)=n^b+b$ for another constant $b>0$.
By Proposition \ref{key-proposition}(1), there exists an NTM $N$ recognizing the language $K=\{1^{f(n)-|x|-1}\# x\mid n\in\nat, |x|\leq f(n), x\in L(N_n)\}$ in polynomial time using space $O(\log{e(|w|)})$ together with advice of size $O(e(|w|))$ for inputs $w$, where $e(|w|)=g\circ f^{-1}(|w|)$. 
The membership of $K$ to $\mathrm{NSC}^k/2^{O(\log^kn)}$ follows easily.

By our assumption, we obtain $K\in\mathrm{SC}^k/2^{O(\log^kn)}$, which implies the existence of a DTM $M$ recognizing $K$ in time $|w|^c+c$ using $c \log^k|w|+c$ space with the help of advice of size $2^{c\floors{\log^k|w|}+c}$ for inputs $w$, where $c$ is an appropriate positive constant.
It then follows by Proposition \ref{key-proposition}(2) that there is a family $\{M_n\}_{n\in\nat}$ of $e'(n)$-size 2dfa's solving $\LL$, where $e'(n)=2^{O(s\circ f(n))}+ O(d\circ f(n))$.
It is possible to prove that $\LL$ belongs to $\ptime2^{\log^k\twod}/\poly$, as requested.


(Only if--part)
Assuming that $\ptime2^{\log^k\twod}/\poly = \ptime2^{\log^k\twon}/\poly$, let us consider any language $K$ in $\mathrm{NSC}^k$ over an alphabet $\Sigma$ and take an NTM $N$ running in $|x|^a+a$ time using at most $a\log^k{|x|}+a$ space for $K$, where $a$ is an appropriate constant in $\nat^{+}$. For convenience, we set $s(n)=a \log^kn+a$ for all $n\in\nat$ and also set $f(n)=n$ for all $n\in\nat$. By Proposition \ref{second-key-proposition}(1) with the help of Lemma \ref{two-choice-change}, there exists a family $\{N_n\}_{n\in\nat}$ of $e(n)$-size $O(1)$-choice 2nfa's solving $\LL=\{(L_n^{(+)},L_n^{(-)})\}_{n\in\nat}$, where $L_n^{(+)}=\{x\in\Sigma^* \mid |x|=f(n), x\in K\}$, $L_n^{(-)}=\{x\in\Sigma^* \mid |x|=f(n), x\notin K\}$, and $e(n)=2^{O(s\circ f(n))}$,
which equals $2^{O(a\log^kn+a)} = 2^{O(\log^kn)}$.
It thus follows that the family $\LL=\{(L_n^{(+)},L_n^{(-)})\}_{n\in\nat}$ belongs to $\ptime2^{\log^k\twon}/\poly$.

Our assumption therefore ensures that $\LL$ falls in $\ptime2^{\log^k\twod}/\poly$ by
a suitable family $\{M_n\}_{n\in\nat}$ of $2^{O(s(n))}$-size 2dfa's solving $\LL$ in time polynomial in $(n,|x|)$.
Proposition \ref{second-key-proposition}(2) then yields a DTM $M$ that recognizes $K$ in polynomial time using $O(\log^k{e'(n)})$ space with advice of size $O(e'(|x|)^2)$, where $e'(|x|)=g(|x|)=2^{a\log^k|x|+a}=2^{O(\log^kn)}$. These facts imply that $K$ belongs to $\mathrm{SC}^k/2^{O(\log^kn)}$.
\end{proofsketch}

In sharp contrast to Proposition \ref{log-k-twod-SC}, in the case of runtime constraint, the complexity classes  $2^{\log^k\twod}/\poly$ and  $2^{\log^k\twon}/\poly$ do not differ. This situation is similar to Proposition \ref{twonpd-with-poly}.

\begin{proposition}\label{coincide-log-k-twod}
For each integer $k\geq2$, $2^{\log^k\twod}/\poly$ coincides with $2^{\log^k\twon}/\poly$.
\end{proposition}

The proof of this proposition is, in essence, similar to the proof of Proposition \ref{log-k-twod-SC}.

\begin{proposition}\label{separation-ALL}
For any integer $k\geq2$, $\mathrm{ALL}/\poly \neq 2^{\log^k\twodpd}/\poly$.
\end{proposition}

\section{Promise Problem Families of Exponential Ceilings}\label{sec:exp-ceilings}

\begin{figure}[t]
\centering
\resizebox*{!}{5.2cm}{\includegraphics*{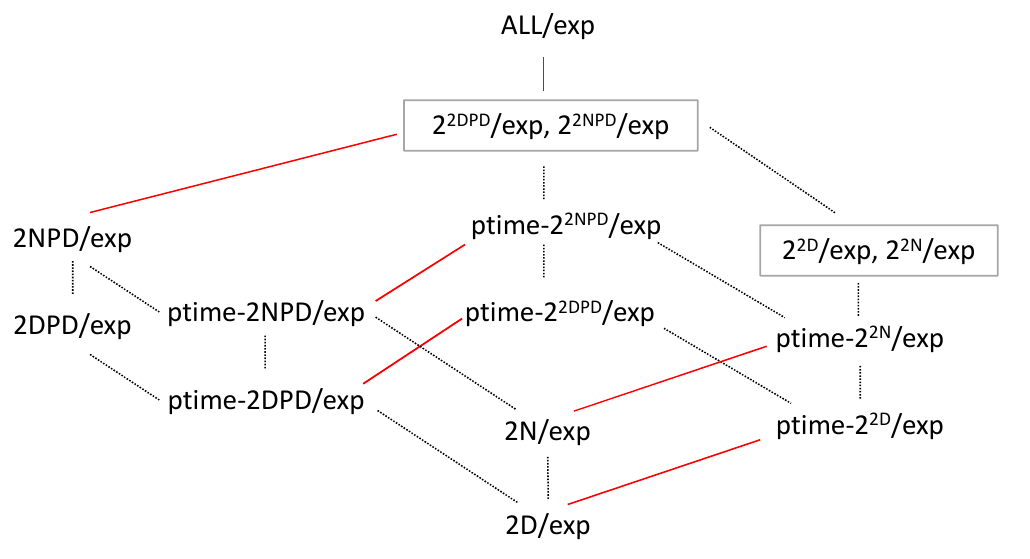}}
\caption{Inclusion relationships among nonuniform complexity classes with exponential ceilings except for the separation result of Theorem \ref{separation-exp-ceiling}. Solid red lines indicate \emph{proper} inclusions.}\label{fig:hierarchy-exp}
\end{figure}

We turn our attention to families of promise problems with exponential ceilings. An early result along this line of study includes Kapoutsis's  result \cite{Kap14} on a close connection between the $\twod/\mathrm{exp} =? \twon/\mathrm{exp}$ question and the $\mathrm{NSPACE}(O(\log\log{n})) \subseteq? \mathrm{DSPACE}(O(\log\log{n}))/\polylog$ question.
In this section, we further explore class relationships among nonuniform complexity classes.
The next claims follow by the proper use of Propositions \ref{key-proposition}-\ref{second-key-proposition} in a way similar to proving Proposition \ref{log-k-twod-SC}.
Most of our results of this section are illustrated in Fig.~\ref{fig:hierarchy-exp}.

\begin{proposition}\label{exp-twod-twon}
(1) $2^{\twod}/\mathrm{exp} = 2^{\twon}/\mathrm{exp}$.

(2) $\ptime2^{\twod}/\mathrm{exp} = \ptime2^{\twon}/\mathrm{exp}$ if and only if $\mathrm{NSC}\subseteq \mathrm{SC}/2^{\mathrm{polylog}}$.
\end{proposition}

\begin{proofsketch}
(1)  It is clear that $2^{\twod}/\mathrm{exp} \subseteq 2^{\twon}/\mathrm{exp}$. In what follows, we intend to prove the opposite inclusion.
Let $\LL=\{(L_n^{(+)},L_n^{(-)})\}_{n\in\nat}$ denote any family in $2^{\twon}/\mathrm{exp}$ over alphabet $\Sigma$. Let $c$ denote a constant in $\nat^{+}$ satisfying $L_n^{(+)}\cup L_n^{(-)}\subseteq \Sigma^{\leq 2^{n^c+c}}$ for all $n\in\nat$.
By setting $f(n)=2^{n^c+c}$, we define
$K=\{1^{f(n)-|x|-1}\# x\mid n\in\nat, |x|\leq f(n), x\in L(N_n)\}$.
By Proposition \ref{key-proposition}(1), this language $K$ falls in $\mathrm{NSPACE}(O(\log^{O(1)}n))/2^{\polylog}$.
Since  $\mathrm{DSPACE}(O(\log^{O(1)}n))/2^{\mathrm{polylog}}$ equals $\mathrm{NSPACE}(O(\log^{O(1)}n))/2^{\mathrm{polylog}}$ without any assumption,  $K$ also belongs to $\mathrm{DSPACE}(O(\log^{O(1)}n))/2^{\polylog}$. By Proposition \ref{key-proposition}(2), the family $\LL$ belongs to $2^{\twod}/\mathrm{exp}$.

(2) The proof of this statement is similar in essence to (1). The runtime restriction of underlying 2nfa's and 2dfa's affects the runtime of an NTM and a DTM.
\end{proofsketch}

In certain circumstances, it is possible to separate two complexity classes without relying on any unproven assumptions. The following separation is one of them.

\begin{proposition}\label{ALL-separation-exp}
$\mathrm{ALL}/\mathrm{exp} \neq 2^{\twodpd}/\mathrm{exp}$.
\end{proposition}

A basic idea of the proof of this proposition is the same as that of the proof of Proposition \ref{separation-ALL}.

The following is an exponential-ceiling version of Theorem \ref{ptime-log-k-subset}.

\begin{theorem}\label{separation-exp-ceiling}
$\ptime2^{\twod}/\mathrm{exp} \nsubseteq \twonpd/\mathrm{exp}$.
\end{theorem}

Theorem \ref{separation-exp-ceiling} leads to a variety of consequences.
Here, we list only two of them without proofs.

\begin{corollary}
(1) $\ptime2^{\twon}/\mathrm{exp} \nsubseteq \twon/\mathrm{exp}$.
(2) $\ptime2^{\twod}/\mathrm{exp} \nsubseteq \twod/\mathrm{exp}$.
\end{corollary}

\begin{corollary}
(1) $\ptime2^{\twonpd}/\mathrm{exp} \nsubseteq \ptime\twonpd/\mathrm{exp}$.
(2) $\ptime2^{\twodpd}/\mathrm{exp} \nsubseteq \ptime\twodpd/\mathrm{exp}$.
\end{corollary}


Cook's argument in \cite{Coo71} does not apply to complexity classes whose space is bounded below $\log{n}$.  From this fact and the proposition below, we may not be able to determine whether or not  $\twodpd/\mathrm{exp}$ and $\twonpd/\mathrm{exp}$ coincide.  This is in sharp contrast with Proposition \ref{twonpd-with-poly}.

\begin{proposition}\label{two-exp-result}
(1) It follows that $\twodpd/\mathrm{exp} = \twonpd/\mathrm{exp}$ if and only if $\dauxpdaspti{O(\log\log{n})}{\infty}/\polylog = \nauxpdaspti{O(\log\log{n})}{\infty}/\polylog$.

(2) It follows that $\ptime\twodpd/\mathrm{exp} = \ptime\twonpd/\mathrm{exp}$ if and only if $\dauxpdaspti{O(\log\log{n})}{n^{O(1)}}/\polylog = \nauxpdaspti{O(\log\log{n})}{n^{O(1)}}/\polylog$.
\end{proposition}

The proof of this proposition is by the use of Propositions \ref{key-proposition}--\ref{second-key-proposition}.

To compensate Proposition \ref{two-exp-result}, we intend to demonstrate the collapse of $2^{\twonpd}/\mathrm{exp}$ down to $2^{\twodpd}/\mathrm{exp}$.  This is a pushdown-automata counterpart of Proposition \ref{exp-twod-twon}(1).

\begin{proposition}
$2^{\twodpd}/\mathrm{exp} = 2^{\twonpd}/\mathrm{exp}$.
\end{proposition}

\begin{yproof}
Let $\LL =\{(L_n^{(+)},L_n^{(-)})\}_{n\in\nat}$ be any family in $2^{\twonpd}/\mathrm{exp}$ over alphabet $\Sigma$.  Take  a family $\{N_n\}_{n\in\nat}$ of $2^{n^a+a}$-size 2npda's that solves $\LL$ for a suitable constant $a>0$. Take another constant $c>0$ for which $\LL$ is $2^{n^c+c}$-ceiling. Now, let $f(n)=2^{n^c+c}$ and $g(n)=2^{n^a+a}$.
By Proposition \ref{key-proposition}(1), since $e(n) = g\circ f^{-1}(n) = 2^{n^{O(1)}}$, we obtain $K\in \nauxpdaspti{O(\log^{O(1)}n)}{\infty}/2^{\polylog}$. By an advice version of Cook's result \cite{Coo71}, $K$ also belongs to $\dauxpdaspti{O(\log^{O(1)}n)}{\infty}/2^{\polylog}$.
Consider a DTM that recognizes $K$ using space $\log^kn+k$ with advice of $2^{\log^bn+b}$ size. Let $s(n)=\log^kn+k$ and $d(n)=2^{\log^b+b}$.
By Proposition \ref{key-proposition}(2), $\LL$ belongs to $2^{\twodpd}/\mathrm{exp}$ because $e'(n)=2^{O(s\circ f(n))}+O(d\circ f(n)) = 2^{n^{O(1)}}$.
\end{yproof}

\section{Complexity of One-Way Pushdown Automata Families}\label{sec:two-way-aux}

Throughout Sections \ref{sec:poly-ceilings}--\ref{sec:exp-ceilings}, we have explored relationships among nonuniform complexity classes induced by families of two-way finite and pushdown automata. Except for a few cases, such as Theorems \ref{ptime-log-k-subset} and \ref{separation-exp-ceiling}, it seems difficult to prove their collapse/separations without any unproven assumptions. When we discuss similar situations associated with ``one-way'' automata families, it is sometimes (not always) possible to demonstrate collapse/separations among complexity classes induced by those automata families.
In this section, we intend to  discuss such separations.
Later, we will briefly discuss how polynomial ceilings affect collapse/separations of those complexity classes.

We first remark that  $\oned\neq \onen$ \cite{SS78}, $\twobp\subseteq 2^{\oned}$, $2^{\oned}\nsubseteq \onep$, and $\onep\subsetneqq 2^{\onep}$ \cite{Yam19a} and that  $\oneu\nsubseteq \onedpd$ and $\mathrm{1t}\onedpd\nsubseteq \onep$  \cite{Yam23}, where ``1t'' refers to ``one turn.''   As new additions to those known results,
we present the following separation results.

\begin{theorem}\label{onedpd-bound}
(1) $\mathrm{1t}\onedpd\nsubseteq 2^{\onen}$. (2) $\co2^{\linoner}\nsubseteq \onenpd$ (and thus $2^{\linonep}\nsubseteq \onenpd$).
\end{theorem}

Theorem \ref{onedpd-bound}(1)  should be compared to  Theorem \ref{Cook-proof}. Moreover,
by analyzing the proof of Proposition \ref{ALL-equivalent}, it follows that $\mathrm{ALL}/\poly = 2^{\onedpd}/\poly = 2^{\onenpd}/\poly$.


\begin{proofsketchof}{Theorem \ref{onedpd-bound}}
(1) We first define $L_n^{(+)}=\{a^mb^m\mid m\geq n\}$ and $L_n^{(-)}= a^*b^* - L_n^{(+)}$ for each index $n\in\nat$ and then consider the family $\LL=\{(L_n^{(+)},L_n^{(-)})\}_{n\in\nat}$. It can be shown easily that $\LL$ is in $\mathrm{1t}\onedpd$.
Next, we wish to show that $\LL\notin 2^{\onen}$. Assume otherwise and take a family $\MM=\{M_n\}_{n\in\nat}$ of exponential-size 1nfa's solving $\LL$.
There is  a polynomial $p$ satisfying  $|Q_n|\leq 2^{p(n)}$ for all $n\in\nat$. We then apply a (standard) pumping lemma for 1nfa families
to $\MM$.

(2)
As an example language $\LL'=\{(L_n^{+)},L_n^{(-)})\}_{n\in\nat}$, we define $L_n^{(+)}=\{a^mb^mc^m \mid m\leq 2^{2^n}\}$ and $L_n^{(-)}=a^*b^*c^* -  L_n^{(+)}$ for each index $n\in\nat$. It is not difficult to prove that $\LL'$ belongs to $\co2^{\linoner}$.

To show that $\LL'\notin\onenpd$, we assume otherwise and take a family $\{M_n\}_{n\in\nat}$ of polynomial-size 1npda's that solves $\LL'$. We apply  to $L(M_n)$ a pumping lemma for context-free languages.
Let $s(n)$ denote the stack-state complexity of $M_n$.
It is important to remark that the pumping lemma constant for $L(M_n)$ is at most $s(n)^2\cdot 2^{s(n)\log{s(n)}+1}< 2^{2^{n}}$.
For any input $w$ of the form $a^mb^mc^m$ with $m=2^{2^{n}}$, $w$ can be factorized into $xyzuv$  so that  $(y,u)$ is an iterative pair of $w$ for $L(M_n)$. However, this violates the definition of $L_n^{(+)}$.

The second part of (2) comes from the fact that $2^{\linoner}\cup \co2^{\linoner} \subseteq 2^{\linonep}$.
\end{proofsketchof}

\begin{proposition}\label{various-separation}
$2^{\log^k\onenpd} \nsubseteq \onenpd$.
\end{proposition}

\begin{yproof}
We note that if $2^{\log^k\onen}\subseteq \onenpd$ then $2^{\log^k\onen}/\poly \subseteq \onenpd/\poly$ follows.
By the nondeterministic version of Corollaries \ref{twod-poly-twodpd} and \ref{log-k-2D-poly}, we obtain $\leq^{\twof}_{m}\!(2^{\log^k\onen}/\poly) = 2^{\log^k\twon}/\poly$ and  $\leq^{\twof}_{m}\!(\onenpd/\poly) = \twonpd/\poly$. Hence, we conclude that $2^{\log^k\onen}/\poly \subseteq \onenpd/\poly$ implies $\ptime2^{\log^k\twon}/\poly \subseteq \twonpd/\poly$. However, this contradicts Theorem \ref{ptime-log-k-subset}.
\end{yproof}


It is important to remark that all separations in Propositions \ref{onedpd-bound} and \ref{various-separation} are proven in the case of no ceiling bound.
When promise problem families are limited to, e.g.,  having polynomial ceilings, separations of some of the classes of one-way automata families become difficult as the following proposition suggests.

\begin{proposition}\label{one-way-case-collapse}
(1) If $\oned/\poly = \onen/\poly$, then $\dl/\poly = \nl/\poly$.

(2)  If $\onedpd/\poly = \onenpd/\poly$, then $\logdcfl/\poly = \logcfl/\poly$.
\end{proposition}

\begin{yproof}
(1)
Assume that $\oned/\poly = \onen/\poly$. This implies that $\leq^{\twof}_{m}\!(\oned/\poly) = \:  \leq^{\twof}_{m}\!(\onen/\poly)$. By Corollary \ref{twod-poly-twodpd}(1), we conclude that $\twod/\poly = \twon/\poly$. As shown in \cite{Kap14}, this is logically equivalent to $\dl/\poly=\nl/\poly$.

(2)
A similar argument works by an application of Corollary \ref{twod-poly-twodpd}(2). The conclusion comes from the fact that $\ptime\twodpd/\poly = \ptime\twonpd/\poly$ iff $\logdcfl/\poly = \logcfl/\poly$ \cite{Yam21}.
\end{yproof}



\end{document}